\documentclass[sigconf,nonacm]{acmart}

\usepackage{booktabs} 

\usepackage{listings}
\usepackage{xcolor}

\definecolor{codegreen}{rgb}{0,0.6,0}
\definecolor{codegray}{rgb}{0.5,0.5,0.5}
\definecolor{codepurple}{rgb}{0.58,0,0.82}
\definecolor{backcolour}{rgb}{0.95,0.95,0.92}

\lstdefinestyle{mystyle}{
    backgroundcolor=\color{backcolour},   
    commentstyle=\color{codegreen},
    keywordstyle=\color{magenta},
    numberstyle=\tiny\color{codegray},
    stringstyle=\color{codepurple},
    basicstyle=\ttfamily\tiny,
    breakatwhitespace=false,         
    breaklines=true,                 
    captionpos=b,                    
    keepspaces=true,                 
    numbers=left,                    
    numbersep=5pt,                  
    showspaces=false,                
    showstringspaces=false,
    showtabs=false,                  
    tabsize=2
}

\usepackage[ruled]{algorithm2e} 

\SetAlFnt{\small}
\SetAlCapFnt{\small}
\SetAlCapNameFnt{\small}
\SetAlCapHSkip{0pt}

\begin{document}
\title{One algebra for all : Geometric Algebra methods for neurosymbolic XR scene authoring, animation and neural rendering}

\begin{teaserfigure}
\centering
\includegraphics[width=0.98\linewidth]{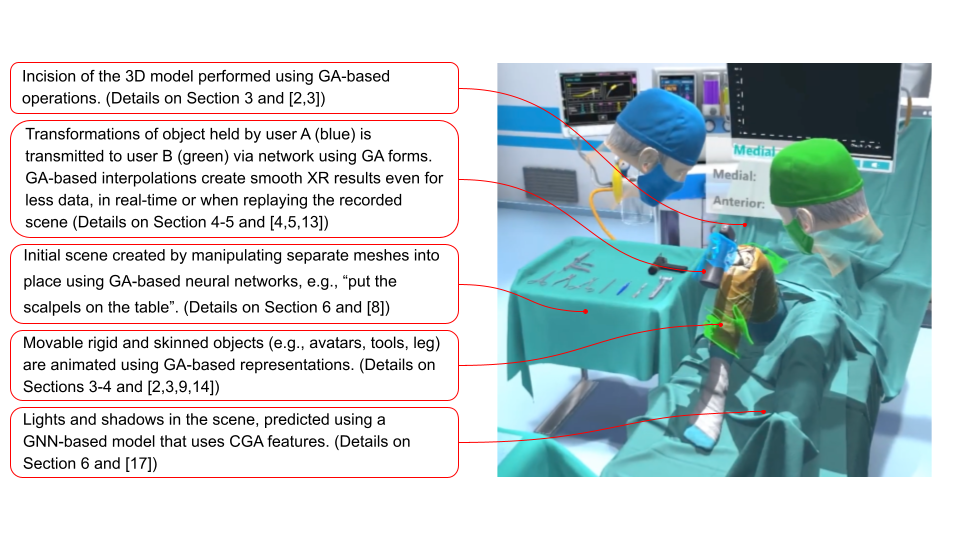}
\caption{The GA-Powered Graphics Pipeline: Geometric Algebra as one coherent framework unifying and enhancing generative AI, character animation, neural rendering, and networked XR.}
\label{fig:pipeline_overview}
\end{teaserfigure}

\author{Manos Kamarianakis}
\orcid{0000-0001-6577-0354}
\affiliation{%
  \institution{FORTH - ICS, University of Crete, ORamaVR}
  \city{}
  \country{}
}
\author{Antonis Protopsaltis}
\orcid{0000-0002-5670-1151}
\affiliation{%
  \institution{University of Western Macedonia, ORamaVR}
  \city{}
  \country{}
}
\author{George Papagiannakis}
\orcid{0000-0002-2977-9850}
\affiliation{%
  \institution{FORTH - ICS / University of Crete / ORamaVR}
  \city{}
  \country{}
}
\renewcommand{\shortauthors}{Kamarianakis, Protopsalts, Papagiannakis}

\begin{abstract}
This position paper delves into the transformative role of Geometric 
Algebra (GA) in advancing specific areas of Computer Graphics (CG) and 
Extended Reality (XR), particularly in character animation, rendering, 
rigging, neural rendering, and generative AI-driven scene editing. Common 
CG algorithms require handling rotations, translations, and dilations 
(uniform scalings) in operations such as object rendering, rigged model 
animation, soft-body deformation, and XR simulations. Traditional 
representation forms—such as matrices, quaternions, and vectors—often 
introduce limitations in precision and performance. Recent breakthroughs 
in the use of GA suggest it can significantly enhance these processes by 
encapsulating geometric forms and transformations into uniform algebraic 
expressions, which maintain critical geometric properties throughout 
multi-step transformations. Furthermore, we explore how GA can serve as a 
unifying mathematical substrate for neurosymbolic XR scene authoring, 
bridging learned neural representations and explicit geometric reasoning. 
This paper outlines how GA-based approaches can improve the fidelity of 
rigged character animations, enhance soft-body simulations, streamline 
real-time rendering, and optimize neural and generative AI scene editing. 
GA offers a coherent and efficient framework for these processes, 
resulting in superior visual outcomes and computational efficiency, 
particularly in XR environments.
\end{abstract}



%
%

\keywords{Geometric Algebra, Real-time Rendering, Soft-Body Simulation, Generative Scene Editing}

\maketitle

\section{Introduction}
\label{sec:introduction}

The field of Computer Graphics (CG) has historically relied on a fragmented 
collection of mathematical algebras and geometrical systems to represent and manipulate 3D objects, 
scenes and transformations. Developers routinely switch between vectors for position, matrices for 
linear transformations, and quaternions or dual-quaternions for handling 
rotations to avoid issues like gimbal lock. While each of these formalisms is 
powerful in its specific domain, their coexistence within a single pipeline 
necessitates constant, computationally expensive conversions. This 
fragmentation not only introduces performance overhead and potential for 
numerical instability but also creates a conceptual disconnect 
and lack of intuition: the underlying 
geometry is often obscured by the disparate algebraic representations. For example, Euler angles provide an intuitive description of rotations, whereas quaternions, though more robust computationally, obscure the geometric intuition behind them.

This paper posits that Geometric Algebra (GA)---a Clifford Algebra over the real numbers \cite{hestenes2012clifford}---provides a comprehensive and unified mathematical language capable of 
streamlining the entire computer graphics pipeline—from modeling and animation 
to simulation, rendering, and interactive XR manipulation. GA offers a single, 
intuitive framework where geometric entities (points, lines, planes, spheres) 
and transformations (rotations, translations, dilations) are represented as 
uniform algebraic objects called multivectors. Our position is that by adopting 
GA, we can develop algorithms that are not only more computationally efficient 
and numerically robust, but also more geometrically intuitive. This unified 
approach preserves the intrinsic meaning of geometric operations throughout 
complex computations, leading to higher-fidelity results in character 
simulation, interactive deformation, and generative scene manipulation.

Our research program has demonstrated the practical benefits of this position 
across several critical areas of CG and Extended Reality (XR). We have 
developed novel algorithms for real-time mesh deformation, cutting, and tearing 
that leverage the expressive power of GA to handle complex topological changes 
\cite{J1, C4-B1}. Our award-winning work in character animation has shown that 
GA-based skinning and interpolation methods can reduce data transmission in 
networked environments while improving animation quality \cite{C6-B2, C15}. These 
advances culminated in the introduction of \emph{GA-Unity}, the first Unity 
package that seamlessly integrates GA principles into collaborative, networked 
applications. GA-Unity extends the benefits of GA into practical 3D scene 
management and rendering, offering an accessible interface for real-time 
multivector operations that enhance both runtime efficiency and bandwidth usage 
in multiplayer scenarios. We have further validated these methods by 
integrating them into a modern XR authoring platform, enabling high-performance, 
artifact-free interactions in demanding applications such as surgical 
simulations \cite{J7, C3}. More recently, we are extending this framework to the 
domain of generative AI, exploring how GA can provide a structure-preserving 
latent space for intuitive, instruction-based scene manipulation with Large 
Language Models (LLMs) \cite{P2}.

This paper is structured to highlight this unified perspective across the 
entire graphics pipeline. Section~\ref{sec:GAFramework} introduces the core 
principles of Geometric Algebra as a unified framework for geometric modeling. 
Subsequent sections describe its applications to mesh manipulation, character 
animation \cite{papaefthymiou2016inclusive}, and networked XR environments, including the GA-Unity framework as a 
key integration layer for real-time deployment. 
Figure \ref{fig:pipeline_overview} provides a visual summary of this vision, illustrating how GA can serve as a unified substrate for the entire modern graphics pipeline, a concept we will detail in the following sections.
Finally, we synthesize these directions to 
outline the emerging vision of GA as a foundational language for geometry, 
simulation, and intelligence in computer graphics.


\section{Geometric Algebra as a Unified Framework for Computer Graphics}
\label{sec:GAFramework}
At its core, GA extends familiar vector algebra 
with a new, invertible product: the geometric product. For any two 
basis vectors $\mathbf{u}$ and $\mathbf{v}$, their geometric product 
$\mathbf{uv}$ is defined as the sum of their inner (dot) and outer 
(wedge) products:
$$ \mathbf{uv} = \mathbf{u} \cdot \mathbf{v} + \mathbf{u} \wedge \mathbf{v} $$
The scalar part, $\mathbf{u} \cdot \mathbf{v}$, captures metric 
information such as length and angle, while the new entity, 
$\mathbf{u} \wedge \mathbf{v}$, is a \emph{bivector}. A bivector 
represents the oriented plane segment swept out by the two vectors. 
This simple addition is profound. By combining scalars, vectors, 
bivectors, and higher-grade elements (trivectors for volume, etc.) 
into a single algebraic structure, GA allows us to represent geometric 
objects and the operations acting upon them in a unified manner.

Building upon this foundation, the \emph{3D Euclidean Geometric Algebra (EGA)}, 
denoted as $\mathbb{R}^3$, provides the fundamental model for representing 
points, lines, planes, and rotations in ordinary three-dimensional space.  
It consists of three orthogonal basis vectors $\mathbf{e}_1$, 
$\mathbf{e}_2$, and $\mathbf{e}_3$, together with all their outer 
products, yielding \(2^3 = 8\) basis elements in total.  
EGA thus spans scalars, vectors, bivectors, and a trivector, enabling a 
compact and intuitive encoding of geometric entities and operations.  
Rotations, for example, are naturally represented by \emph{rotors}---
multivectors that generalize quaternion-based rotations without losing geometric clarity. 
More interested readers may consult \cite{dorst2009geometric,hildenbrand2013foundations,survey2024Eckhard} for comprehensive treatments of Geometric Algebra and its applications.

Extending further, the \emph{Conformal Geometric Algebra (CGA)} has emerged 
as a particularly powerful model for computer graphics. CGA, the geometric algebra of 
$\mathbb{R}^{4,1,0}$, unifies rigid-body transformations (rotations and 
translations) as operations within a single algebraic structure, and 
represents points, spheres, and planes in 3D space as 5D vectors (null 
vectors of the algebra). The notation $\mathbb{R}^{4,1,0}$ indicates four 
positive, one negative, and zero null dimensions in the metric signature.  
The resulting 5D conformal model arises by extending the 3D Euclidean basis 
with two additional vectors, $\mathbf{e}_0$ and $\mathbf{e}_\infty$, 
corresponding respectively to the origin and the point at infinity.  
Although the conformal basis is five-dimensional, the complete algebra 
contains \(2^5 = 32\) elements, as all possible outer-product combinations 
of the basis vectors are included. Rotations, translations, and dilations 
are all represented by special multivectors (called rotors, translators, 
and dilators) that act on geometric objects through the ``sandwich'' product:
$$ \text{Object}_{\text{transformed}} = 
M \cdot \text{Object}_{\text{original}} \cdot M^{-1} $$
where $M$ is the multivector representing the transformation.

This structure allows GA to generalize and subsume other algebraic 
systems used in computer graphics. Quaternions, for instance, are 
isomorphic to the bivectors in 3D Euclidean GA, and dual-quaternions 
find their natural home within the even subalgebra of 3D Projective 
Geometric Algebra (PGA) or CGA~\cite{J1}. This means that any operation 
performed with quaternions or dual-quaternions has a direct, and often 
more intuitive, counterpart in GA. 

The fundamental property that makes GA so powerful is that 
\emph{geometry is preserved through algebraic manipulation}. When we 
compose transformations by multiplying their corresponding multivectors 
($M_{\text{composite}} = M_2 M_1$), the resulting multivector 
$M_{\text{composite}}$ correctly represents the composite geometric 
transformation. When we interpolate between two transformations by 
linearly interpolating their multivectors, the result is a geometrically 
meaningful intermediate transformation~\cite{J6, C6-B2}. This is not 
true for matrices where, for example, interpolating two rotation 
matrices does not generally yield another rotation matrix. This 
property simplifies CG pipelines immensely, eliminating the need for 
special interpolation schemes like SLERP for quaternions and removing 
the conceptual clutter of managing multiple, incompatible mathematical 
representations.

Beyond its theoretical elegance, GA’s unifying capability now extends 
into practical deployment. The recently introduced \emph{GA-Unity} 
framework \cite{C15} operationalizes these principles within the Unity game engine, 
offering a modular environment where multivector-based transformations 
are computed natively. 
The relationship among these representation forms and their seamless transmutation into GA equivalents is illustrated in Figure~\ref{fig:ga_unity}, where GA-Unity facilitates conversion between conventional transformation representations (vectors, matrices, quaternions) and their geometric-algebraic counterparts within modern game engines.
By enabling automatic conversion between GA 
objects and standard engine constructs, GA-Unity bridges the gap 
between mathematical abstraction and real-time implementation. It 
supports efficient interpolation, deformation, and scene manipulation 
in collaborative, networked applications, demonstrating that the 
benefits of GA are not confined to theoretical rigor but translate 
directly into measurable performance gains and development efficiency 
within contemporary 3D engines.

\section{Applications in Geometry Processing and Mesh Manipulation}

One of the most compelling demonstrations of Geometric Algebra's 
power is in the domain of real-time geometry processing and 3D 
mesh manipulation. Traditional methods for simulating deformable 
objects or performing interactive mesh modifications like cutting, 
tearing, and drilling often rely on complex data structures 
(e.g., tetrahedral meshes) and computationally intensive methods 
like the Finite Element Method (FEM). These approaches frequently 
require heavy pre-processing and struggle to maintain interactive 
frame rates, especially in immersive XR applications.

Our research has shown that CGA provides a remarkably compact and 
efficient framework for these tasks \cite{J1, C4-B1}. By representing 
vertices, planes, and spheres as uniform multivectors, we can 
express complex geometric predicates and operations as simple 
algebraic manipulations. For example, the intersection of a 
cutting plane and a mesh edge can be determined directly through 
GA products, simplifying the logic required for real-time cutting.

In our work \cite{J1}, we introduced a unified geometric algorithm to 
cut, tear, and drill deformable, rigged models. This framework 
leverages CGA to perform these operations on-the-fly, even after 
the model has been animated, while robustly maintaining the 
deformation topology. Unlike previous methods, our approach 
requires minimal pre-processing and is ``GA-ready,'' meaning all 
sub-predicates are implemented in terms of multivector operations. 
This unified approach yields algorithms that are not only compact 
but also achieve excellent numerical stability and topological 
robustness. For instance, during a cut operation, our method 
creates new vertices along the intersection and instantly computes 
their skinning weights using barycentric coordinates. This allows 
the newly separated mesh parts to be further deformed and 
animated smoothly and without artifacts. This ability to perform 
dynamic topological changes on animated characters is crucial for 
realistic surgical XR simulations and other interactive training 
applications \cite{J7}.

A core benefit of this GA-native approach is its radical 
simplification of geometric constraints. The algebra itself 
encodes the geometric relationships, freeing developers from 
managing complex conditional logic and data structures required 
by traditional methods. This results in cleaner, more maintainable 
code and delivers real-time performance, essential for demanding 
interactive XR environments.

\section{Rigged Character Animation and Scene-Level Transformations}

\begin{figure*}
    \centering
    \includegraphics[width=0.98\linewidth]{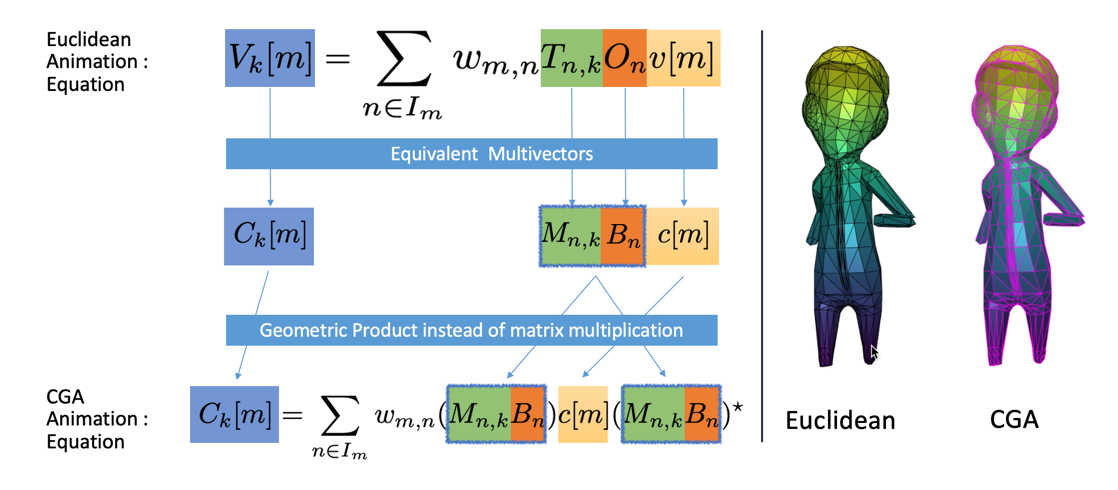}
    \caption{The ``original'' euclidean animation equation involves vectors and matrices can be transformed to an equivalent one that only involves multivectors \cite{C4-B1, J1}. Results for keyframes are identical, with interpolated frames showing minor deviations (see  Figure~\ref{fig:vr_recorder}). \textit{Notation:} $V_{k}[m]$ is the position vector of the position of the $m$-th vertex at animation time $k$ (in homogeneous coordinates), $T_{n,k}$ is the 4x4 matrix storing the transformation of the $n$-th bone at animation time $k$ and $O_{n}$ is the offset matrix corresponding to the $n$-th bone. $w_{m,n}$ denotes the weight of the $n$-th bone on the $m$-th vertex, $I_m$ is the set of indices of bones that affect the $m$-th vertex and $v[m]$ is the original position of the $m$-th vertex. $C_k[m], M_{n,k}, B_n$ and $c[m]$ are the equivalent multivector forms of $V_k[m], T_{n,k}, B_n$ and $v[m]$.}
    \label{fig:animation_euc_vs_cga}
\end{figure*}

Animating rigged characters is a cornerstone of modern CG. The 
standard method, Linear Blend Skinning (LBS), is prone to 
well-known artifacts like collapsing joints (``candy-wrapper'' 
effect). In contrast, Dual-Quaternion Skinning (DQS) largely solves 
many of these issues, achieving superior volume preservation and 
more realistic deformations by blending rigid-body transformations 
(rotations and translations) directly, rather than matrices \cite{papaefthymiou2016inclusive,papagiannakis2013geometric}. 
Despite its advantages, DQS does not elegantly handle non-uniform 
scaling, a common requirement for stylized animation.

Geometric Algebra offers a natural and more powerful extension to 
these techniques. As dual-quaternions can be represented within GA, 
any DQS algorithm has a direct equivalent in a GA framework. More 
importantly, GA provides a unified way to handle all affine 
transformations, including scaling. In CGA, rotations, translations, 
and dilations (i.e., uniform scalings) are all represented by multivectors (rotors, 
translators, and dilators) that can be composed via the geometric 
product.

Our research has leveraged this unified handling of transformations 
to improve both the fidelity and efficiency of deformable character 
animation. In our award-winning work \cite{C6-B2, C15}, we developed a 
methodology for rigged character animation and deformation entirely 
within CGA. The standard skinning equation, which blends 
transformations on a per-vertex basis, is translated into its 
multivector equivalent. This allows us to apply a blend of rotations, 
translations, and dilations in a single, coherent framework, using a 
simple sandwich product to deform the vertices.
$$ C_{k}[m]=\sum_{n\in I_{m}}w_{m,n}(M_{n,k}B_{n})c[m](M_{n,k}B_{n})^{-1} $$
Here, the final vertex position $C_{k}[m]$ is a weighted sum of 
the vertex $c[m]$ being transformed by the composite multivector 
$(M_{n,k}B_{n})$ for each influencing bone (see Figure~\ref{fig:animation_euc_vs_cga}).

This approach not only simplifies the underlying mathematics but 
also yields tangible performance and quality benefits. We have 
demonstrated that linear interpolation of GA multivectors (motors) 
produces smooth, artifact-free intermediate animation frames. This 
capability is especially valuable for generating keyframes on-the-fly, 
enabling high-fidelity motion that directly enhances the perceived 
realism of the animation.

\begin{figure*}
    \centering
    \includegraphics[height = 130pt, ]{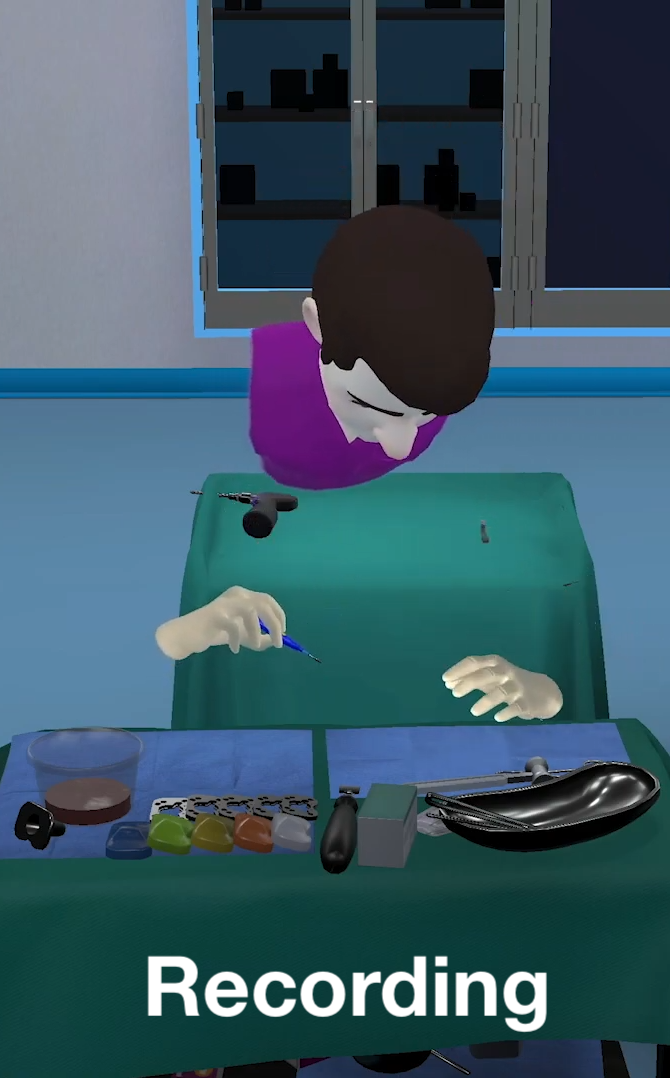}
    \includegraphics[height = 130pt, ]{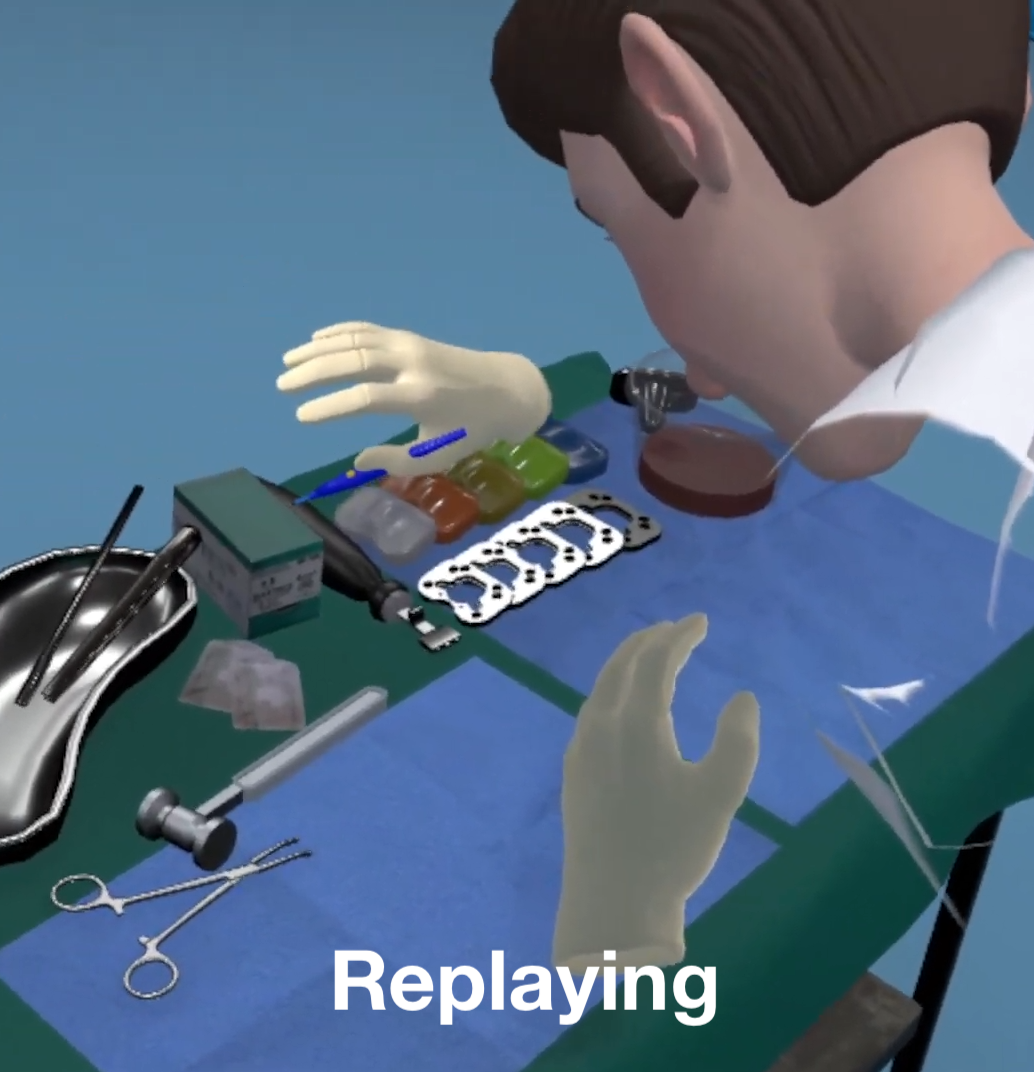}
    \includegraphics[height = 130pt, ]{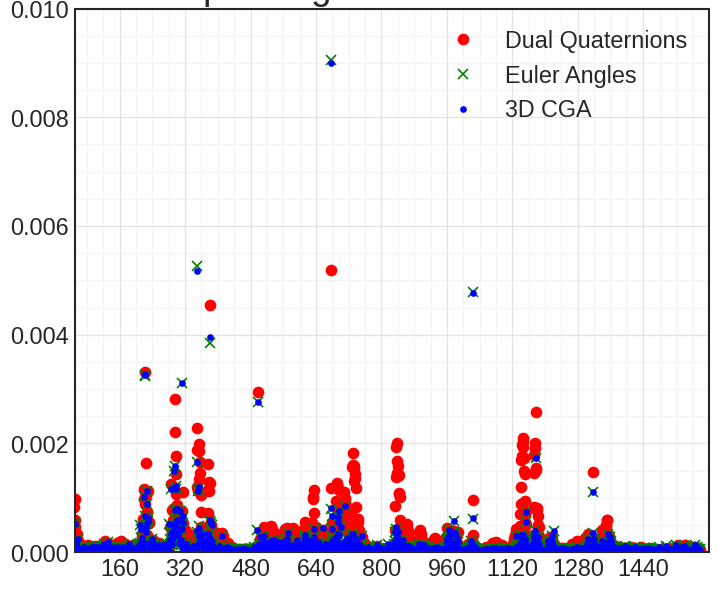}
    \caption{GA for Accurate Psychomotor Action Replay: Our VR Record and Replay system (left, center) enables high-fidelity playback of user actions for analysis or training \cite{C9}. (Right) The system's accuracy is built on GA, which provides a more robust transformation representation with significantly lower relative error compared to traditional quaternions. }
    \label{fig:vr_recorder}
\end{figure*}

Beyond traditional animation, this same GA-based representation 
has proven highly effective for capturing, encoding, and replaying 
full-body motion in immersive environments. In our VR Recorder and 
Replay system \cite{J6}, user movements are captured as GA 
multivectors, allowing each recorded session to be compactly stored 
and later replayed with high geometric fidelity. By leveraging GA’s 
compact and expressive representation, we can significantly reduce 
the amount of stored motion data, while maintaining interpolation 
accuracy comparable to standard representations such 
as Euler angles, vectors, or even dual-quaternions (see Figure~\ref{fig:vr_recorder}). This capacity to 
capture and reproduce motion precisely further validates GA’s 
versatility as a unified mathematical foundation for  the graphics pipeline.

\section{Networked and Collaborative XR Environments}

The demands of real-time, multi-user XR applications place a 
significant strain on network bandwidth. In a collaborative virtual 
environment, the position, orientation, and scale of every 
user-controlled object and avatar must be continuously synchronized 
across all clients. Transmitting full transformation matrices (16 
floats) for every update is inefficient and can lead to high 
latency, which breaks the sense of shared presence and immersion.

This representational efficiency provides a clear and measurable advantage in networked environments, acritical factor for real-time XR. A rigid-body transformation (rotation and translation) represented using a GA multivector (a ``motor'' in Projective or Conformal GA) is far more compact than a 
4x4 matrix. For example, a motor in 3D PGA requires only 8 floats, 
equivalent to a dual-quaternion. Leveraging this compactness 
significantly reduces data payload that must be transmitted over the network for each transformation update.

\begin{figure*}
    \centering
    \includegraphics[height = 94pt, ]{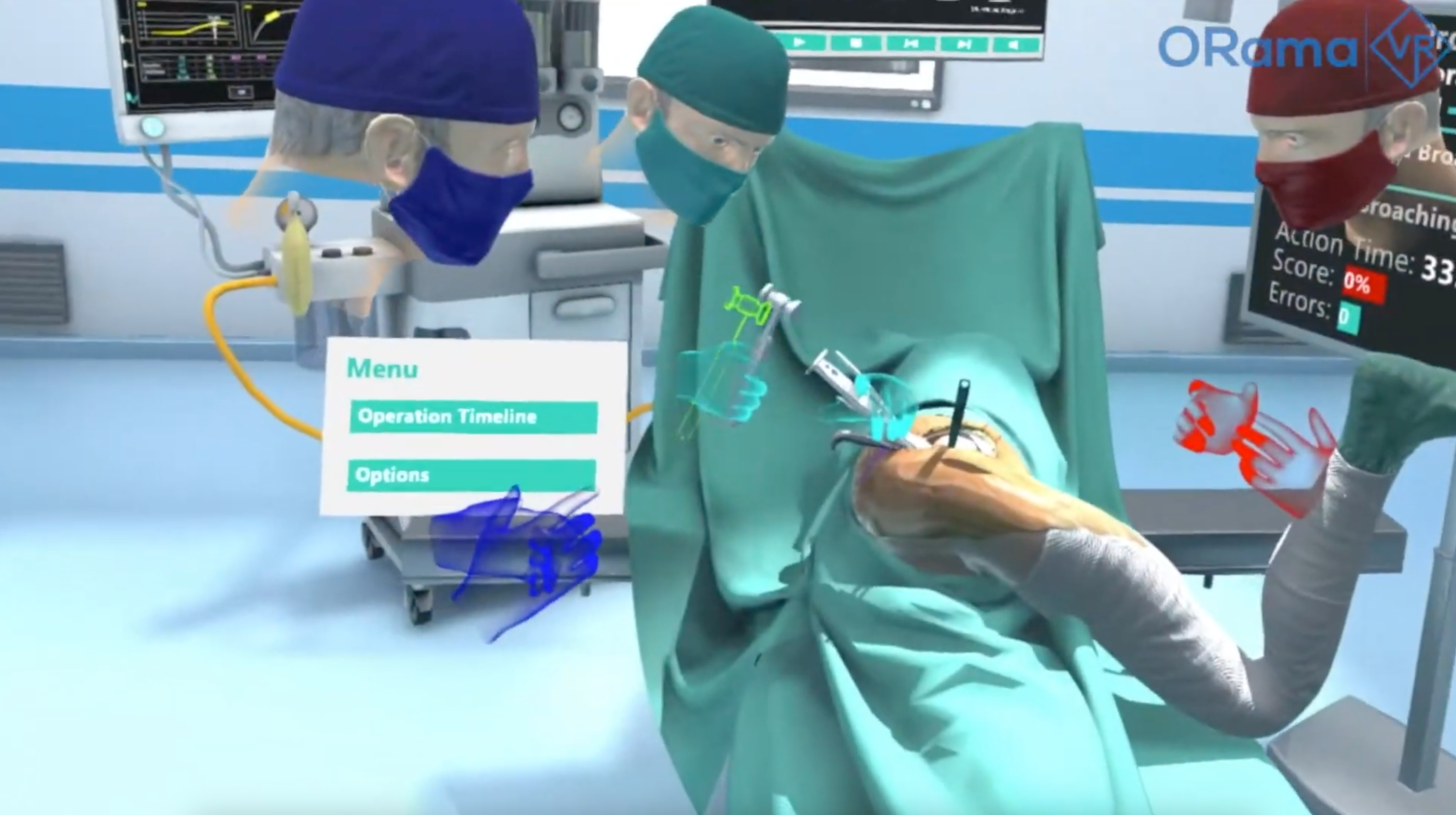}
    \includegraphics[height = 94pt, ]{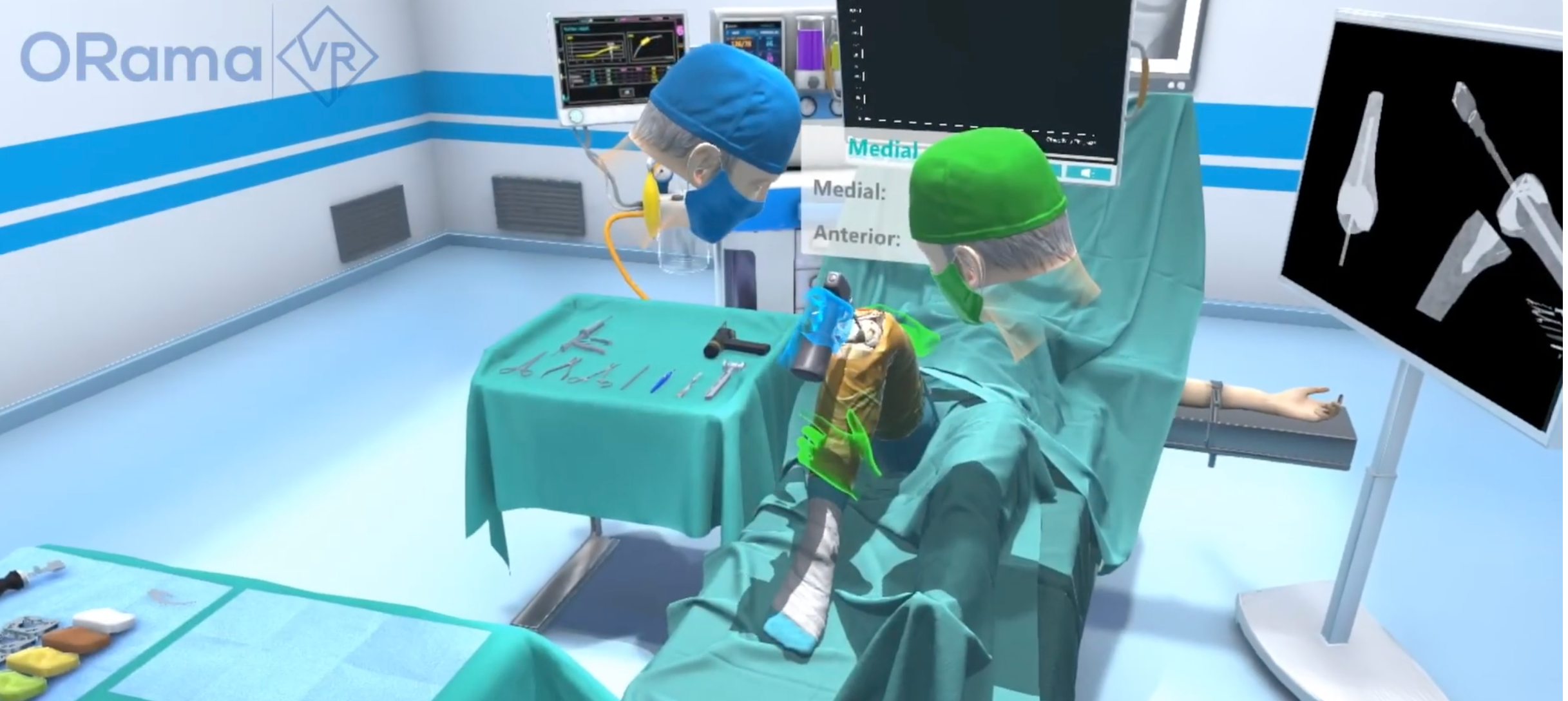}
    \caption{Robust Real-Time Surgery Training in XR: Our GA framework powers high-fidelity collaborative medical training, enabling dynamic, artifact-free mesh cutting and deformation on deformable patient models \cite{C3,J7}.}
    \label{fig:networked}
\end{figure*}

Our research, detailed in \cite{J6, C6-B2}, validates this approach experimentally. We implemented a shared VR system where transformation data for objects and hand-controllers are encoded as GA multivectors before transmission. Critically, the 
client side receives these compact multivectors and used them to locally  interpolate intermediate frames. This GA-based pipeline reduced the required network bandwidth by up to 58\% compared to traditional vectors and quaternion-based methds, while achieving an equivalent or superior Quality of Experience (QoE).

This efficiency comes from the power of multivector interpolation. 
Because interpolating between two GA motors yields a geometrically 
meaningful results, we can send fewer keyframe updates per second and 
rely on the client to generate smooth, jitter-less motion locally. 
This is particularly effective in bandwidth-restricted cases; for instance, 
our GA-based method can achieve a visual quality 30 updates/sec with a traditional approach using only 20 updates/sec \cite{C6-B2}. 
This reduction in data traffic is critical for scaling collaborative 
XR experiences to include more users and dynamic objects, establishing
it as a key enabling technology for multiplayer gaming, 
remote collaboration, and shared educational simulations (see Figure \ref{fig:networked}), work that was recognized through a poster presentation at SIGGRAPH 2022 \cite{C9}.

\section{From Geometry to Intelligence: GA in Neural and Generative Graphics}

The recent explosion of generative AI and Large Language Models 
(LLMs) has opened new frontiers for content creation and scene 
manipulation. The ability to edit a 3D scene using natural language 
instructions---``move the chair to the right of the table''---is a 
paradigm shift for artists, designers, and everyday users. However, a 
fundamental challenge remains: bridging the gap between the semantic, 
often ambiguous, nature of human language and the precise, 
mathematical descriptions required by a graphics engine. Current 
methods often rely on machine learning models that operate on 
unstructured representations like pixels or voxels (e.g., NeRFs CITE NEEDED), 
which makes precise, object-level manipulation difficult.

We propose that Geometric Algebra can serve as the ideal 
structure-preserving latent space for AI-driven graphics. 
Because GA multivectors encode both geometric entities and their 
spatial relationships within a single algebraic framework, they 
provide a perfect symbolic language for an AI to reason about a 3D 
scene. An LLM's task is simplified from generating raw coordinate 
data to generating the correct algebraic operators (GA multivectors) 
that achieve the desired transformation.

Our current research on LLM-based scene manipulation, ``Shenlong'' 
\cite{P2}, demostrates this GA-AI integration. The system leverages an LLM 
to interpret a user's natural language command, with a crucial distinction: the LLM's role is not to generate code or vertex positions, but to translate user intent into a precise CGA expression. For example, 
the instruction ``place object A on top of object B'' causes the LLM to 
construct a CGA expression. This epression, when executed, resolves
to a CGA translator multivector that computes the necessary 
displacement based on the objects' bounding boxes. This final CGA motor
is then applied directly to the object's transformation 
component in the scene graph.

\begin{figure*}
    \centering
    \includegraphics[width=0.98\linewidth]{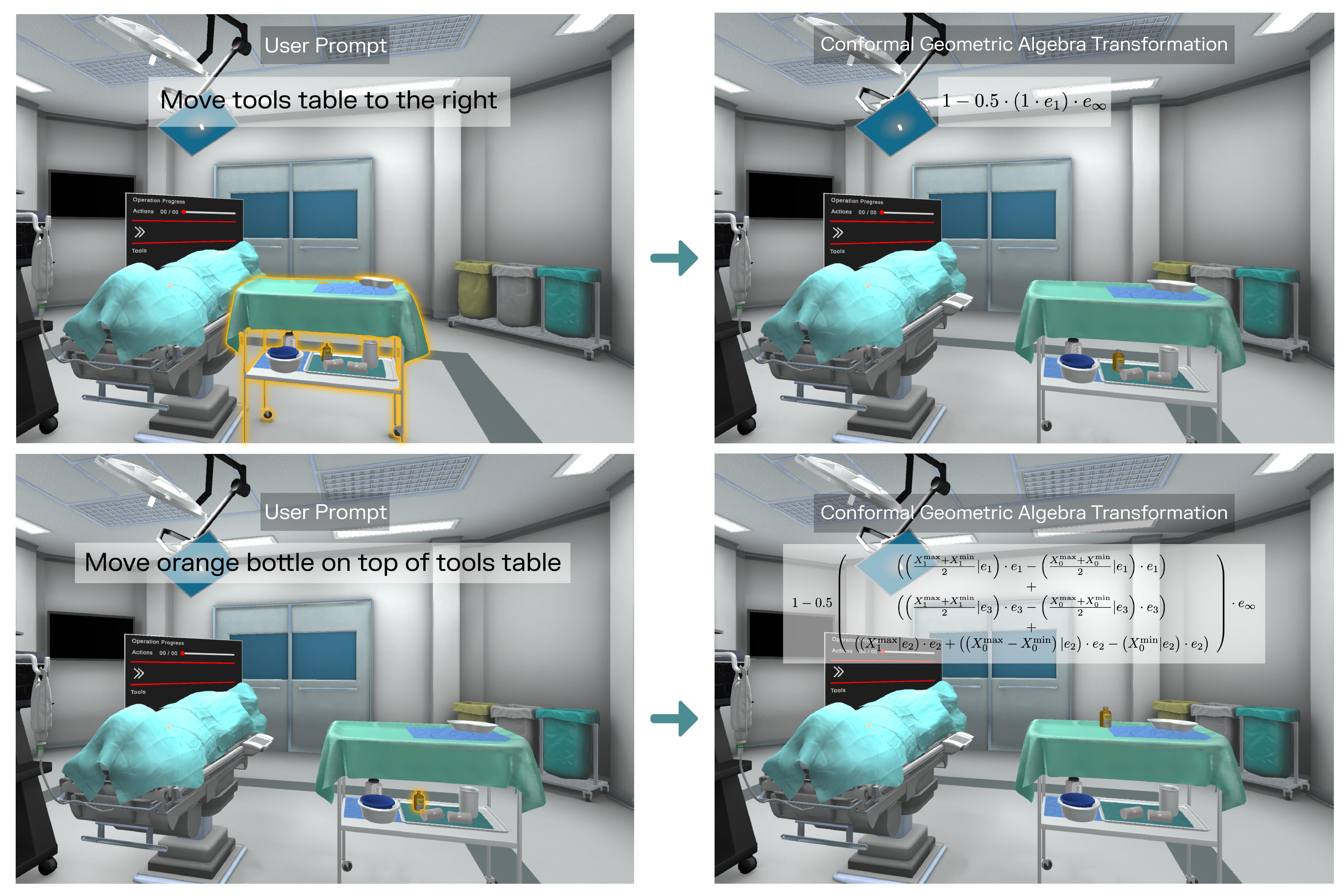}
    \caption{AI-Driven Editing with Geometric Precision: The ``shenlong'' system, where a LLM translates ambiguous user intent ("on top of") into a mathematically precise CGA expression for robust 3D manipulation \cite{P2}.}
    \label{fig:shenlong}
\end{figure*}

This GA-centric approach has several advantages. First, it leverages the 
zero-shot reasoning capabilities of pre-trained LLMs, eliminating the need for massive, scene-specific training datasets. Second, it 
ensures geometric precision, as the final transformation is a 
mathematically exact GA operation, not a probabilistic approximation. Our preliminary results (see Figure \ref{fig:shenlong}) validate this method, showing that 
it significantly outperforms baselines by reducing 
LLM response times and boosting success rates on complex spatial 
queries.

\begin{figure*}
    \centering
    \includegraphics[width=0.98\linewidth]{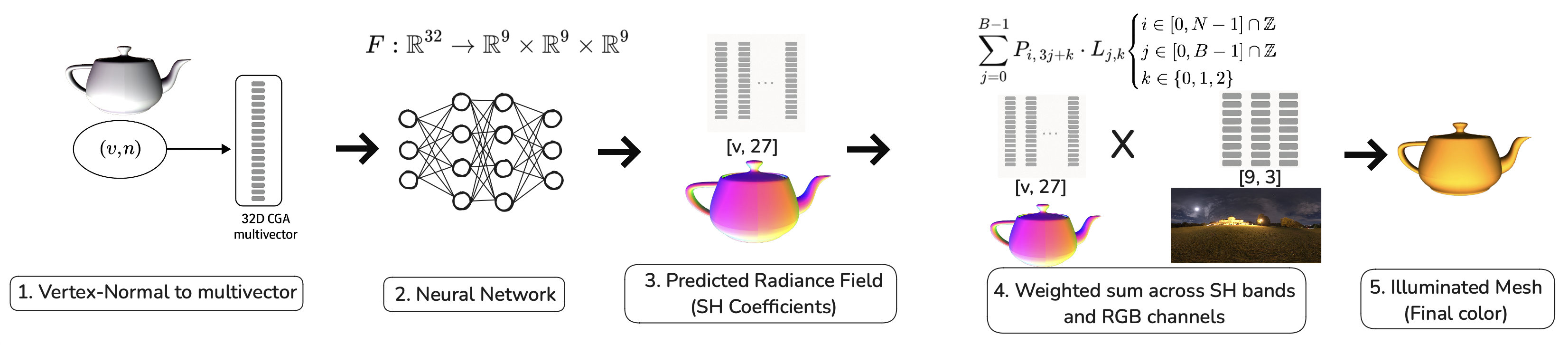}
    \caption{A Geometrically-Aware Neural Architecture: The NeuralGASh pipeline consumes CGA multivectors—which compactly encode vertex position and normal data—to directly predict Spherical Harmonic (SH) coefficients for light and shadow, eliminating traditional pre-computation steps \cite{geronikolakis2025neural}.}
    \label{fig:neural_gash_pipeline}
\end{figure*}

\begin{figure*}
    \centering
    \includegraphics[width=0.49\linewidth]{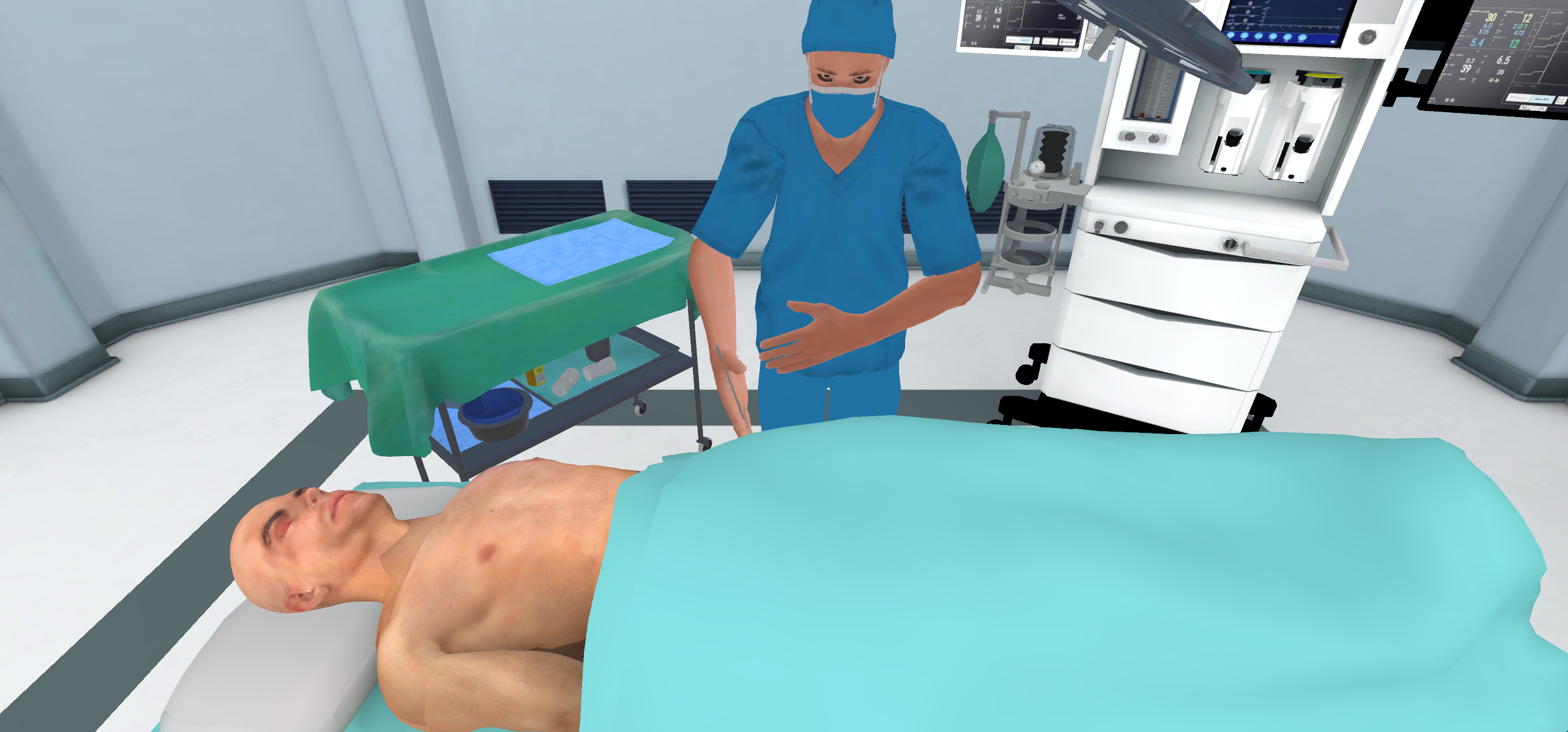}
    \includegraphics[width=0.49\linewidth]{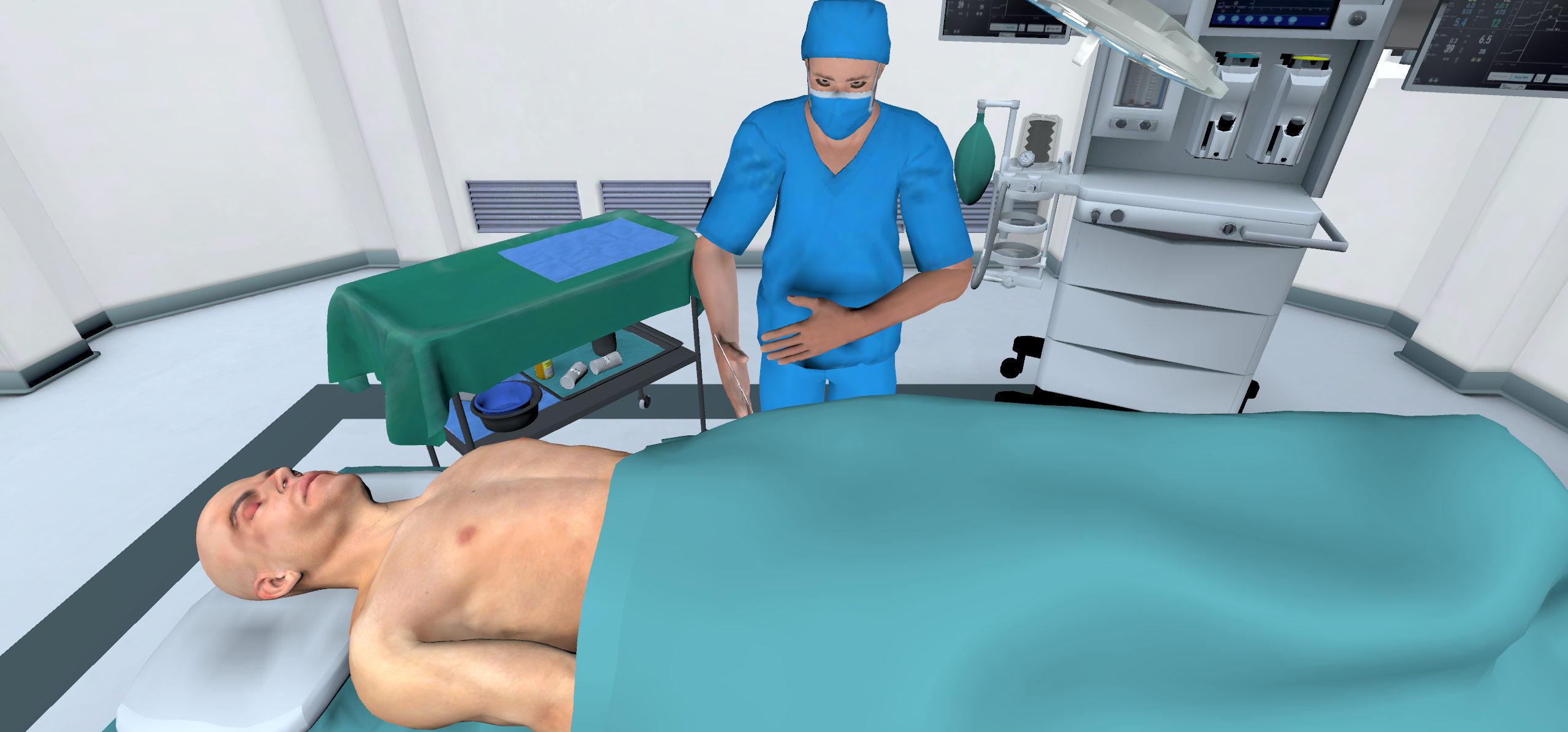}
    \caption{Enhanced Realism with GA-Powered Neural Rendering: (Left) Standard Unity engine lighting on a dynamic character. (Right) The same scene rendered with our "NeuralGASh" pipeline, which achieves superior, dynamic lighting and shadow fidelity without pre-computation \cite{geronikolakis2025neural}. }
    \label{fig:neural_gash}
\end{figure*}

\begin{figure*}
    \centering
    \includegraphics[width=0.98\linewidth]{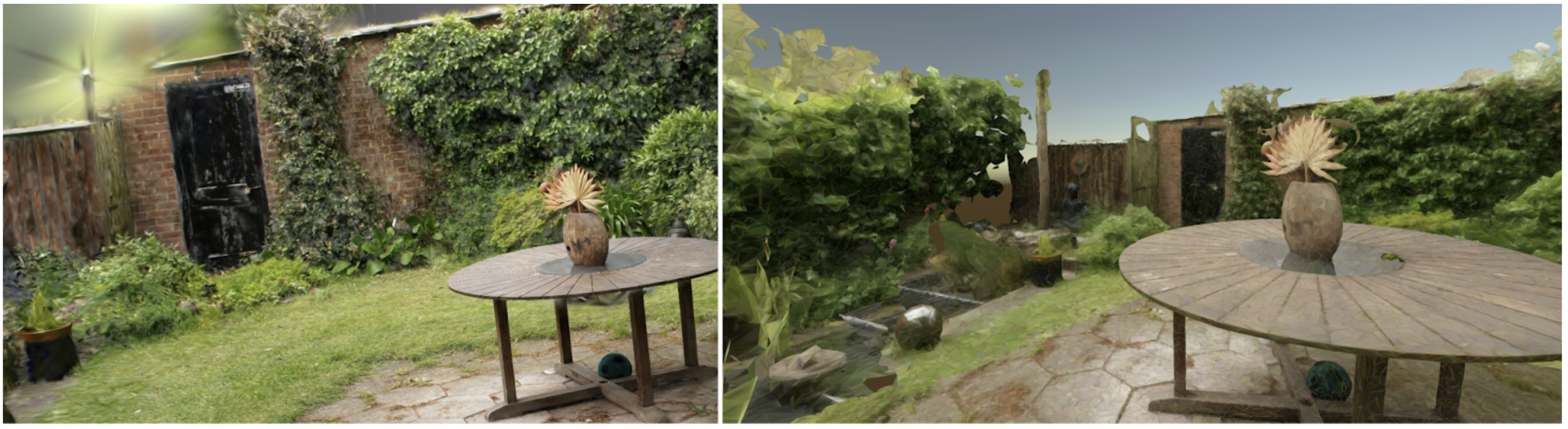}
    \caption{GA-Powered Shading for 3D Gaussian Splats: Our NeuralGASh pipeline demonstrating its robustness and flexibility by providing real-time, dynamic lighting and shadows for modern scene representations like 3D Gaussian Splatting, not just traditional meshes \cite{geronikolakis2025neural}.}
    \label{fig:neuralgash_GaussianSplats}
\end{figure*}

This GA-AI synergy extends beyond high-level logic and into the core rendering pipeline. Our recent work on ``Neural-GASh'' [\ref{fig:neural_gash}] introduces a novel real-time shading architecture based on a neural radiance field. This approach, which utilized an initial Multi-Layer Perceptron (MLP) model (see Figure \ref{fig:neural_gash_pipeline}), outperforms traditional Precomputed Radiance Transfer (PRT) methods. The key advantage is that our model is trained to directly consume CGA representations of vertex positions and normals, completely eliminating the expensive, static pre-computation step required by PRT. This enables high-fidelity, dynamic shading of fully animated meshes in real-time (see Figure \ref{fig:neural_gash}). The framework's robustness is further demonstrated by its ability to shade scenes generated via 3D Gaussian Splatting, proving its flexibility beyond traditional mesh-based geometry (see Figure \ref{fig:neuralgash_GaussianSplats}). Furthermore, preliminary results from a new GNN-based extension of this method show even greater fidelity, as the GNN architecture better captures mesh topology to yield superior results in complex self-shadowing scenarios. This architectural progression provides a concrete validation for the new class of geometrically-aware neural architectures we propose.

Looking forward, we envision a new class of GA-embedded neural 
architectures. These models would not merely learn pixel correlations 
but would be structured to inherently understand geometric meaning. By training models 
to operate directly on multivector representations, we can create 
generative-AI systems that respect geometric constraints, understand 
symmetry, and can perform complex spatial reasoning, currently out of reach for purely data-driven approaches. This synergy between the symbolic precision of GA
and the learning power of neural networks defines a compelling new frontier for  
intelligent graphics.

\section{Integration into Modern XR Authoring Platforms}

A significant part of our research has focused on the 
practical validation and integration of our GA-based methods into 
modern XR authoring platforms. By primarily focusing on C\# implementations
within the Unity game engine, we bridge the gap between theoretical 
advancements and production-ready tools. This focus on implementation ensures 
that our algorithms can be readily applied in production environments 
for creating interactive and high-performance applications.

We have successfully integrated our GA algorithms 
for mesh manipulation, character animation, and efficient network 
transmission into a cohesive XR authoring platform \cite{J7}. This 
framework built upon contemporary game engines, directly 
exposes our GA-based functionalities to developers.  
Our `GA-Unity` package, for instance, provides a streamlined workflow 
for developers to represent object transformations as GA 
multivectors. It handles the conversion from standard Unity 
formats (Vector3, Quaternion) to multivectors and provides 
optimized interpolation functions essential for 
networked applications \cite{C15}. Demonstrating significant performance improvements over previous methods, this work was recognized with 
both the Best Paper and the Best Application Award at the CGI 2024 
ENGAGE Workshop.

Our platform has been used to build high-fidelity surgical 
simulations where the real-time, artifact-free deformation, 
cutting, and tearing of soft-body tissues is a critical 
requirement \cite{J1, J7}. In these medical VR applications, the 
robustness and performance of our GA algorithms allow for a level 
of realism and interactivity that would be difficult to achieve 
with traditional methods. Similarly, our framework has been used 
to create collaborative training and educational tools (see Figure \ref{fig:vr_recorder}) where the 
efficient synchronization of actions in a shared virtual 
environment is paramount \cite{C9, C16}.

By packaging these advanced geometric methods into accessible 
tools for platforms like Unity, we are lowering the barrier to 
entry for developers. This simultaneously validates the performance 
and utility of GA in a production context and accelerates the 
dissemination of these powerful techniques into broader industry 
and academic curricula, paving the way for wider adoption.

\section{Unified Perspective and Outlook}
\label{sec:weaving}

The research presented in this paper establishes GA not as a mere mathematical 
abstraction, but as a practical framework capable of unifying and transforming 
many parts of the modern graphics pipeline. Across multiple domains—
mesh manipulation, character animation, XR interaction, and generative 
scene editing—GA provides a single, coherent algebraic framework that 
supersedes the traditional fragmented representations based on vectors, 
matrices, and quaternions.

In mesh manipulation, GA allows us to encode vertices, edges, faces, and 
local transformations as multivectors, enabling real-time operations 
such as deformation, cutting, and tearing while preserving geometric 
coherence. Unlike conventional approaches that treat rotation and 
translation separately, GA unifies these transformations, resulting in 
more robust and numerically stable algorithms. This unified framework 
provides a foundation for techniques previously difficult or impossible 
to express compactly, while simultaneously reducing the 
computational overhead associated with multiple converting between 
different algebraic forms.

\begin{figure*}
    \centering
    \includegraphics[width=0.98\linewidth]{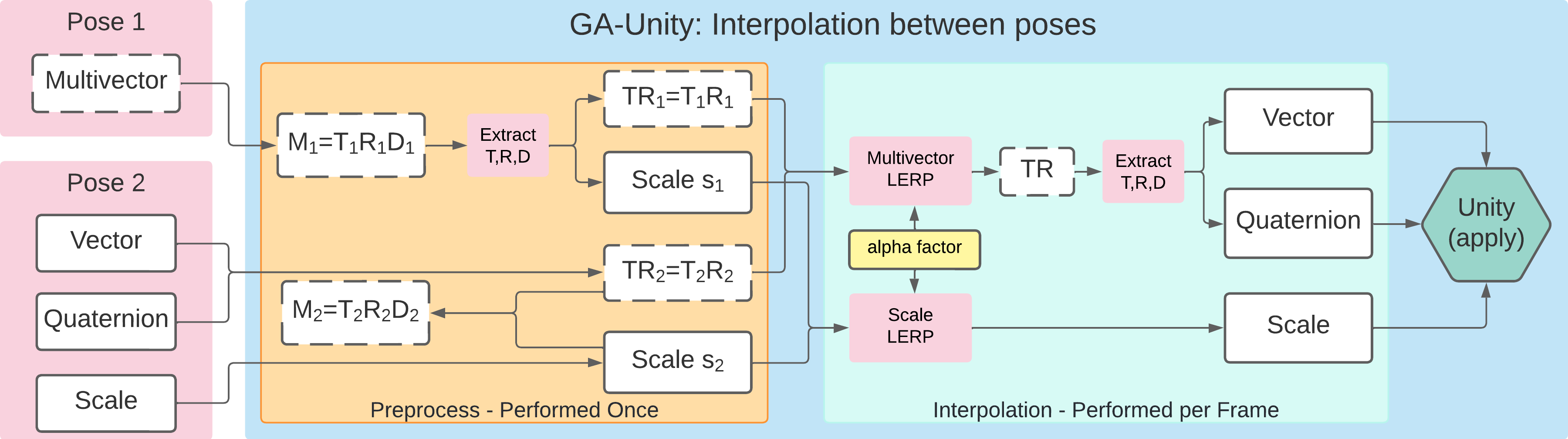}
    \caption{Optimizing Networked XR with GA-Unity: The interpolation engine converts Unity's standard transforms into compact GA multivectors (motors). By interpolating these motors, we generate smooth motion from fewer keyframes, significantly reducing network bandwidth in collaborative applications \cite{C15}.}
    \label{fig:ga_unity}
\end{figure*}

GA extends this unification to character animation, replacing specialized 
schemes like SLERP and dual quaternion interpolation. By representing rotations, 
translations, and interpolations as multivectors, we achieve geometrically 
meaningful blends using a single, consistent algebraic tool. The integration of 
GA into networked systems, as realized in our \emph{GA-Unity} framework, 
directly enables compressed pose transmission, reduces runtime, and 
ensures coherent behavior across distributed simulations. This provides
a tangible improvement over traditional pipelines, demonstrating 
how GA is able to transform not only the mathematical representation but also 
the fundamental system-level architecture for interactive applications.

In XR and collaborative authoring contexts, GA provides a unifying 
language for transformations, deformations, and interactions, ensuring 
that scene updates, manipulations, and physics computations remain 
consistent across local and remote environments. Our GA-Unity framework 
operationalizes these concepts within modern game engines. It handles 
conversions between traditional representations and their GA equivalents (Figure~
\ref{fig:ga_unity}) while allowing developers to leverage 
GA-based operations without modifying existing rendering or simulation 
subsystems. This capability is essential for enabling high-fidelity, multi-user XR experiences where both geometric accuracy and performance are critical.

GA also extends naturally into neural and generative graphics, where 
maintaining geometric fidelity in learned latent spaces is a primary challenge. 
By structuring latent representations as multivectors, we provide
geometrically meaningful foundation for instruction-driven scene editing 
with LLMs \cite{P2}. This integration ensures that semantic 
instructions translate into valid transformations, enabling AI-assisted 
content creation that remains coherent and interpretable, even in complex 
3D environments.

This GA-AI integration is equally transformative in neural rendering. Our "Neural-GASh" architecture \cite{geronikolakis2025neural} demonstrates this by training a neural model to consume CGA representations directly as input. This approach completely eliminates the static pre-computation step of traditional PRT, enabling high-fidelity, dynamic shading of fully animated and deformed meshes. This use of GA as a native input for neural networks provides a concrete validation for a new class of high-performance, geometrically-aware rendering and simulation architectures.

\section{On Practical Adoption and Future Directions}


A critical reader might ask: if GA's benefits are this clear, why has it not already superseded traditional methods? The answer involves a combination of historical inertia, a narrow definition of "performance," and a prior lack of production-ready tools.

Historically, the entire graphics ecosystem—from GPU hardware to rendering APIs—has been aggressively optimized for 4x4 matrix operations. This created a significant performance barrier for alternative algebras. However, this "operation-level" performance view is too narrow for modern, complex applications. As this paper has demonstrated, the system-level benefits of GA are now proving more significant. The massive reduction in network bandwidth (Section 5) and the complete elimination of static pre-computation in neural rendering (Section 6) represent performance gains that far outweigh any micro-level cost of a multivector operation.

The primary barrier, therefore, has been one of usability and tooling. The perceived learning curve of GA and the lack of integration into standard engines have historically prevented widespread adoption. Our work on GA-Unity (Section 7) was developed specifically to solve this problem. By providing an accessible, high-performance bridge that integrates GA-native operations directly into a mainstream engine, we lower the barrier to entry, allowing developers to leverage GA's power without needing to be experts in the underlying algebra. As the demands of XR and generative AI continue to push traditional fragmented tools past their limits, we believe that practical, high-performance frameworks like GA-Unity finally pave the way for GA's wider adoption as the new standard.

Collectively, these contributions establish GA as a comprehensive 
substrate for graphics computation: it not only replaces fragmented 
mathematical tools but also enables new algorithms and tools that exploit 
its algebraic expressiveness. GA-Unity exemplifies this by bridging 
the theoretical advantages of GA with practical implementation in modern 
engines, empowering developers and researchers to create interactive, 
networked, and generative graphics pipelines that were previously 
difficult to realize. Figure~\ref{fig:pipeline_overview} summarizes this 
vision, illustrating how GA permeates the full pipeline—from geometry 
definition to physics-based animation, real-time rendering, XR interaction, 
and AI-driven scene manipulation—providing a single unifying framework 
that elevates both the efficiency and the expressive power of modern 
graphics systems.

In summary, Geometric Algebra is not only an elegant mathematical 
framework but also a practical catalyst for innovation in computer 
graphics and XR. By preserving geometry across transformations, 
reducing redundancy, and enabling coherent multi-domain operations, GA 
augments existing pipelines and facilitates the creation of new techniques 
and tools that fully exploit the potential of a unified algebraic 
representation. This positions GA as a next-generation foundation for 
graphics, interactive systems, and generative environments alike.



While our work demonstrates production-level performance by 
minimizing overhead, the ultimate step is to challenge the 
hardware-level inertia directly. The industry's reliance on 4x4 
matrices is an optimization for a legacy paradigm. The emergence of 
GA-based, matrix-free implementations, such as the one presented in 
\cite{LookMaNoMatrices}, marks the beginning of this transition. In 
that work, GA operations were executed directly within vertex and 
fragment shaders, achieving superior performance compared to 
state-of-the-art matrix-based pipelines. These findings suggest that 
a paradigm shift is both feasible and beneficial. We therefore 
propose that future GPU and CPU architectures should explore 
GA-native hardware acceleration. Just as modern GPUs now include 
``RT Cores'' for ray tracing and ``Tensor Cores'' for AI, a ``GA Core'' 
that natively accelerates multivector operations—such as the 
sandwich product and motor interpolation—could unlock 
orders-of-magnitude performance gains. This would render the entire 
"performance" debate obsolete and position GA as the undisputed 
substrate for all real-time spatial computation.

Finally, the truest adoption will come from standardization. While GA-Unity serves as a robust proof-of-concept, the community must move toward a standardized, open-source, and highly-optimized library for GA computation, analogous to ``GLM'' for linear algebra. This would provide a common foundation for a new generation of fully-unified engines where graphics, computational physics, and AI are not separate subsystems, but rather different applications of the same, single geometric algebra. This 'grand unification' of all spatial computation remains the ultimate promise of GA, a goal that our work demonstrates is not only possible, but practical.


\section*{ACKNOWLEDGMENTS}
This work was partially funded by the Innosuisse Swiss Accelerator (2155012933-OMEN-E), the Horizon Europe Projects INDUX-R (GA 101135556) and FIDAL (GA 101096146).
We would like to acknowledge the foundational contributions of our colleagues Efstratios Geronikolakis, Prodromos Kolyvakis and Nick Lydatakis to the GA research work cited herein.

\bibliographystyle{ACM-Reference-Format}
\bibliography{references}

\end{document}